\documentclass[prb,showpaces,preprintnumbers,amsmath,amssymb,twocolumn,superscriptaddress,longbibliography]{revtex4-1}
\usepackage{graphicx}
\usepackage{dcolumn}
\usepackage{bm}
\usepackage{amssymb}
\usepackage{amsmath}
\usepackage{epstopdf}
\usepackage{booktabs}
\usepackage{threeparttable}
\usepackage[pdfstartview=FitH, CJKbookmarks=true, bookmarksnumbered=true, bookmarksopen=true, colorlinks, pdfborder=001, linkcolor=blue, anchorcolor=blue, citecolor=blue]{hyperref}
\begin{document}
\title{Fully-gapped superconductivity in single crystals of noncentrosymmetric Re$_6$Zr with broken time-reversal symmetry}
\author{G. M. Pang}
\affiliation{Center for Correlated Matter and Department of Physics, Zhejiang University, Hangzhou 310058, China}
\author{Z. Y. Nie}
\affiliation{Center for Correlated Matter and Department of Physics, Zhejiang University, Hangzhou 310058, China}
\author{A. Wang}
\affiliation{Center for Correlated Matter and Department of Physics, Zhejiang University, Hangzhou 310058, China}
\author{D. Singh}
\affiliation{Department of Physics, Indian Institute of Science Education and Research Bhopal, Bhopal 462066, India}
\author{W. Xie}
\affiliation{Center for Correlated Matter and Department of Physics, Zhejiang University, Hangzhou 310058, China}
\author{W. B. Jiang}
\affiliation{Center for Correlated Matter and Department of Physics, Zhejiang University, Hangzhou 310058, China}
\author{Y. Chen}
\affiliation{Center for Correlated Matter and Department of Physics, Zhejiang University, Hangzhou 310058, China}
\author{R. P. Singh}
\affiliation{Department of Physics, Indian Institute of Science Education and Research Bhopal, Bhopal 462066, India}
\author{M. Smidman}
\email{msmidman@zju.edu.cn}
\affiliation{Center for Correlated Matter and Department of Physics, Zhejiang University, Hangzhou 310058, China}
\author{H. Q. Yuan}
\email{hqyuan@zju.edu.cn}
\affiliation{Center for Correlated Matter and Department of Physics, Zhejiang University, Hangzhou 310058, China}
\affiliation{Collaborative Innovation Center of Advanced Microstructures, Nanjing 210093, China}

\date{\today}

\begin{abstract}
The noncentrosymmetric superconductor Re$_6$Zr has attracted much interest due to the observation of broken time-reversal symmetry in the superconducting state. Here we report an investigation of the superconducting gap structure of Re$_6$Zr single crystals by measuring the magnetic penetration depth shift $\Delta\lambda(T)$ and electronic specific heat $C_e(T)$. $\Delta\lambda(T)$ exhibits an exponential temperature dependence behavior for $T~\ll~T_c$, which indicates a fully-open superconducting gap. Our analysis shows  that a single gap $s$-wave model is sufficient to describe both the superfluid density $\rho_s(T)$ and  $C_e(T)$ results, with a fitted gap magnitude larger than the weak coupling BCS value, providing evidence for  fully-gapped superconductivity in Re$_6$Zr with moderate coupling.

\end{abstract}

\maketitle

\section{Introduction}

Noncentrosymmetric  superconductors (NCS) have attracted a great deal of attention due to the influence of antisymmetric spin-orbit coupling (ASOC) on the superconducting properties, which is induced by the antisymmetric potential gradient arising due to broken inversion symmetry \cite{smidman2017superconductivity,BauerNCS}. Strong ASOC may lift the spin degeneracy of the conduction bands, allowing for superconducting states which are an admixture of spin-singlet and spin-triplet pairing, giving rise to a variety of unique properties \cite{smidman2017superconductivity,BauerNCS,gorkov2001superconducting,frigeri2004superconductivity}. Unconventional superconductivity was evidenced in the first heavy fermion NCS  CePt$_3$Si, where a linear temperature dependence of the magnetic penetration depth at low temperatures and a constant Knight shift across $T_c$ were observed\cite{bonalde2005evidence,yogi2006evidence}. While studies of heavy fermion NCS revealed a range of unusual findings, disentangling the role played by broken inversion symmetry from the effects of strong electronic correlations and magnetism is highly challenging, which spurred an interest in looking for weakly correlated  NCS with singlet-triplet mixing. Evidence for such an admixture was demonstrated in Li$_2$(Pd,Pt)$_3$B, where Li$_2$Pd$_3$B has a fully open gap \cite{Takeya2005,Nishiyama2005,yuan2006s}, but the  gap of Li$_2$Pt$_3$B exhibits line nodes \cite{yuan2006s,Nishiyama2007,Takeya2007}. This change from nodeless to nodal superconductivity upon switching Pd for Pt was explained using a model with a mixture of singlet and triplet states, where there is a relative increase in the size of the spin-triplet component of the order parameter, as the ASOC is increased \cite{yuan2006s}. The Knight shift also shows a marked difference between the two compounds, corresponding to a decrease of the spin susceptibility below $T_c$ for  Li$_2$Pd$_3$B, while for Li$_2$Pt$_3$B this remains constant, indicating greater influence of the ASOC on the superconductivity \cite{Nishiyama2005,Nishiyama2007}.

 Subsequently, the order parameters of a wider range of NCS have been studied, where evidence for nodal superconductivity was also found in quasi-one-dimensional K$_2$Cr$_3$As$_3$\cite{pang2015evidence} and the low carrier system YPtBi \cite{YPtBinode}. Meanwhile, although studies of Y$_2$C$_3$ at higher temperatures were accounted for by fully gapped superconductivity \cite{Y2C3NMR,Y2C3HC,Y2C3MuSR}, very low temperature penetration depth measurements  also indicate the presence of a nodal gap structure \cite{chen2011evidence}. However, most NCS have been found to be fully-gapped superconductors, such as BaPtSi$_3$\cite{bauer2009a}, BiPd\cite{Sun2015,Matano2013,jiao2014anisotropic,yan2016nodeless}, PbTaSe$_2$\cite{PbTaSe2SpecH,pang2016nodeless,wang2016nodeless}, La$_7$Ir$_3$ \cite{barker2015unconventional} and Re$_6$Zr\cite{singh2014detection,matano2016full,khan2016complex,Mayoh2017}. In particular, many NCS have been found compatible with single-gap $s$-wave superconductivity, indicating that the size of any triplet component is very small, and the relationship between the ASOC strength and the degree of singlet-triplet pairing is not entirely understood. Moreover, it has been proposed that some NCS can exhibit topological superconductivity. Both theoretical calculations and experimental studies reveal topological surface states in the NCS  PbTaSe$_2$ and BiPd \cite{Bian2016topological,Guan2016,Sun2015,Neupane2016observation}, which may lead to possible Majorana fermions in the vortex cores when these states are sufficiently close to the Fermi level.

Another notable feature of  several NCS  is the breaking of time-reversal symmetry (TRS). While   broken TRS was previously discovered in the unconventional triplet superconductor Sr$_2$RuO$_4$ \cite{luke1998time,xia2006high}, evidence for this phenomenon has also been found in a number of NCS with fully gapped superconducting states, such as LaNiC$_2$\cite{hillier2009evidence}, La$_7$Ir$_3$\cite{barker2015unconventional}, and Re$_6$Zr\cite{singh2014detection}. In the case of LaNiC$_2$, the symmetry analysis shows that for TRS to be broken at $T_c$, the effect of the ASOC on the superconductivity should be weak, indicating a lack of significant mixed parity pairing \cite{quintainilla2010relativistic}. On the other hand,  Re$_6$Zr crystallizes in the cubic $\alpha$-Mn structure, where the presence of three dimensional irreducible representations of the point group potentially allows for TRS breaking with singlet-triplet mixing \cite{singh2014detection}. It is therefore of particular importance to characterize the superconducting order parameter of Re$_6$Zr. Previous measurements of  the gap structure of polycrystalline samples are accounted for by nodeless single gap $s$-wave superconductivity\cite{singh2014detection,matano2016full,khan2016complex,Mayoh2017}, but different conclusions are drawn from recent point contact spectroscopy results from single crystal measurements, which give evidence for multiple gaps \cite{pradnya2017multiband}. As such, it is important to perform further measurements  of Re$_6$Zr single crystals sensitive to low energy excitations to clarify this issue. In this paper, we study the superconducting order parameter of single crystalline  Re$_6$Zr   by measuring the magnetic penetration depth and specific heat of single crystals, which are both consistent with a  single  nodeless isotropic gap with a moderate coupling strength.

\section{Experimental details}

Single crystals of Re$_6$Zr were synthesized using the Czochralski method, as described in Ref.~\onlinecite{pradnya2017multiband}.  Magnetization measurements were performed using  a superconducting quantum interference device (SQUID) magnetometer (MPMS) with both field-cooling (FC) and zero-field-cooling (ZFC),  with a small applied magnetic field of 10~Oe. The electrical resistivity $\rho(T)$ was measured by using a standard four-probe method from room temperature down to 0.3~K in a $^3$He refrigerator. The specific heat $C(T)$ was measured using  a Physical Property Measurement System (PPMS) with a $^3$He insert down to around 0.6~K. The temperature dependence of the magnetic penetration depth shift $\Delta\lambda(T)=\lambda(T)-\lambda(0)$ was measured using a tunnel-diode-oscillator (TDO) technique in a $^3$He cryostat down to  0.35~K. The operating frequency of the TDO setup is about 7~MHz, with a noise level as low as 0.1~Hz. For the TDO measurements, the sample was cut into a regular shape with dimensions of 700$\times$500$\times$200~$\mu$m$^3$ and mounted onto a sapphire rod so that it may be placed into the coil without making contact. The sample experiences a very small ac field of about 20~mOe along the [100] direction  induced by the coil, which is  much smaller than the lower critical field $H_{c1}$, ensuring that the sample remains in the Meissner state, so that the change of the magnetic penetration depth $\Delta\lambda(T)$ is proportional to the frequency change $\Delta f(T)=f(T)-f(0)$. Here $\Delta\lambda(T)=G\Delta f(T)$ , where the calibration constant $G$ is determined by the geometry of the sample and coil \cite{prozorov2000meissner}.

\section{Results and discussion}

\begin{figure}[t]
\begin{center}
  \includegraphics[width=\columnwidth]{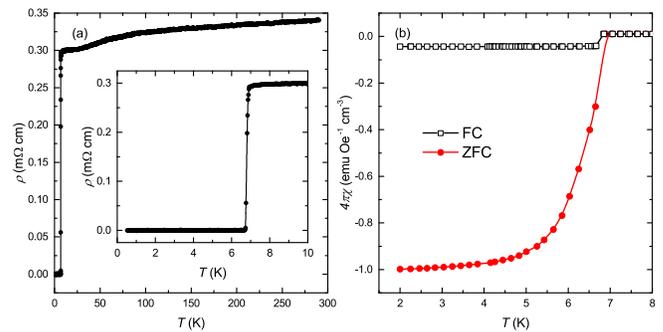}
\end{center}
	\caption{(Color online) Characterization of Re$_6$Zr single crystals showing the temperature dependence of  (a) the electrical resistivity $\rho(T)$ from room temperature down to 0.3~K, and (b) the magnetic susceptibility 4$\pi\chi$ between 2 and 8~K, where both field-cooled (FC) and zero-field cooled curves in an applied field of 10~Oe are displayed. The inset of (a)  shows an enlargement of $\rho(T)$ around $T_c$~=~6.8~K. }
   \label{figure1}
\end{figure}

\begin{figure}[t]
\begin{center}
  \includegraphics[width=\columnwidth]{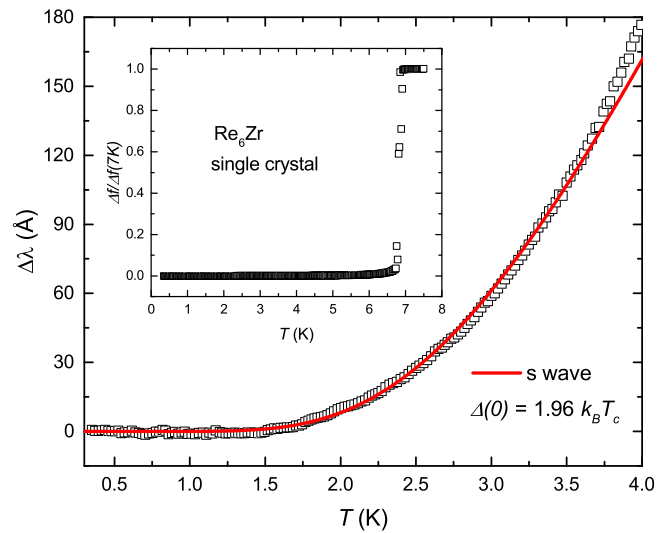}
\end{center}
	\caption{(Color online) Temperature dependence of the penetration depth shift $\Delta\lambda(T)$ at low temperatures, where exponential behavior is observed. The solid red line displays the fitted curve for an $s$-wave model with  $\Delta(0)~=~1.96~k_BT_c$. The inset shows the frequency shift from 7.5~K down to 0.35~K, normalized by the 7~K value, which displays a sharp superconducting transition at $T_c$~=~6.8~K.}
   \label{figure2}
\end{figure}

Single crystals of Re$_6$Zr were characterized by the measurements of the electrical resistivity $\rho(T)$ and magnetic susceptibility $4\pi\chi(T)$, which are displayed in Fig.~\ref{figure1}. $\rho(T)$ exhibits metallic behavior in the normal state with a residual resistivity of $\rho(7K)$~=~0.3~m$\Omega$~cm, just above the superconducting transition. This yields a mean free path of 2.1~nm\cite{orlando1979critical}, with a coherence length of $\xi$~=~5.5~nm deduced from the upper critical field, and a Sommerfeld coefficient of $\gamma_n$~=~27.4~mJ mol$^{-1}$ K$^{-2}$ from our specific heat measurements described below. The calculated mean free path is less than half the coherence length, indicating that the sample is closer to the dirty limit. The inset of Fig.~\ref{figure1} (a) displays an enlargement of $\rho(T)$ at low temperatures, where a sharp superconducting transition occurs at $T_c$~=~6.8~K, with  a transition width  less than 0.2~K. In addition,  the temperature dependence of the magnetic susceptibility in Fig.~\ref{figure1}(b) shows that the zero-field-cooled (ZFC) curve exhibits full diamagnetism, providing evidence for bulk superconductivity in the Re$_6$Zr single crystals.

Figure~\ref{figure2} displays the temperature dependence of the penetration depth shift $\Delta\lambda(T)$ at low temperatures from TDO-based measurements. The frequency change $\Delta f(T)$ is plotted in the inset from about 7.5~K down to the base temperature of 0.35~K, which has been normalized by the normal state value. Here a sharp superconducting transition occurs at around 6.8~K, in line with the  resistivity, and magnetic susceptibility measurements.  $\Delta\lambda(T)$ decreases rapidly upon reducing the temperature, before becoming flat below around 1.5~K, indicating a fully open superconducting gap in Re$_6$Zr. For an $s$-wave superconductor, the change of magnetic penetration depth shows exponentially activated behavior at $T\ll T_c$, as
 \begin{equation}
\Delta\lambda(T)=\lambda(0)[\sqrt{\frac{\pi\Delta(0)}{2k_BT}}\textrm{exp}(-\frac{\Delta(0)}{k_BT})],
\label{equation1}
\end{equation}

\noindent where $\lambda(0)$ and $\Delta(0)$ are the penetration depth and superconducting gap magnitude at zero temperature,  respectively. As shown by the solid red line, our experimental data can be well described by a single $s$-wave gap   with $\lambda(0)$~=~200~nm and $\Delta(0)$~=~1.96~$k_BT_c$. The value of $\lambda(0)$ and the temperature dependence of $\Delta\lambda(T)$ are comparable with those in Ref.~\onlinecite{khan2016complex}, suggesting that both single crystal and polycrystalline samples are consistent with a similar nodeless gap structure. We note that for a superconductor with line nodes in the dirty limit, a quadratic dependence would be expected at low temperatures  instead of a linear $\Delta\lambda(T)$ \cite{Goldenfeld1993}. However, upon analyzing with a power law dependence  $\Delta\lambda(T)\propto T^n$, values of $n$ of 4.4 and 6.1 are obtained  fitting from the base temperature up to 3~K and 2.2~K, respectively. Since these exponents are significantly larger than two, such a dirty nodal scenario can be ruled out.

\begin{figure}[t]
\begin{center}
  \includegraphics[width=\columnwidth]{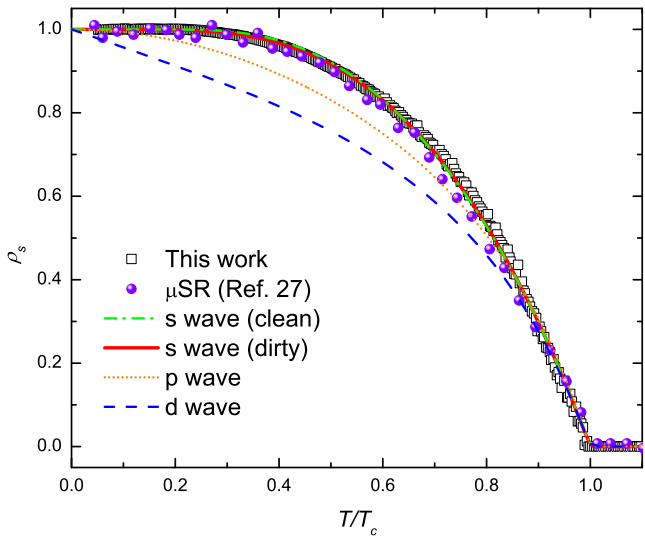}
\end{center}
	\caption{(Color online) Normalized superfluid density $\rho_s(T)$, as a function of the reduced temperature $T/T_c$. The black squares are the   data from this work, while the  dots are  from $\mu$SR results in Ref.~\onlinecite{singh2014detection}. The lines show the superfluid density calculated  from various models of the gap structure.}
   \label{figure3}
\end{figure}

Moreover the magnetic penetration depth was also converted into the normalized superfluid density via  $\rho_s=[\lambda(0)/\lambda(T)]^2$, which is displayed as a function of the reduced temperature $T/T_c$ in Fig.~\ref{figure3}. For a clean superconductor, $\rho_s(T)$ can be calculated using

\begin{equation}
\rho_{\rm s}(T) = 1 + 2 \left\langle\int_{\Delta_k}^{\infty}\frac{E{\rm d}E}{\sqrt{E^2-\Delta_k^2}}\frac{\partial f}{\partial E}\right\rangle_{\rm FS},
\label{equation2}
\end{equation}

\noindent where $f(E, T)=[1+{\rm exp}(E/k_BT)]^{-1}$ is the Fermi-Dirac distribution function and the superconducting gap function is defined as $\Delta_k(T)=\Delta(T)$g$_k$. This  contains an angular dependent part g$_k$ and a temperature dependent component $\Delta(T)$, which can  be approximated as $\Delta(T)$~=~$\Delta(0){\rm tanh}\left\{1.82\left[1.018\left(T_c/T-1\right)\right]^{0.51}\right\}$ \cite{Carrington2003}. Here the zero temperature gap magnitude $\Delta(0)$ is the only fitted parameter. As shown by the green dashed-dotted line in Fig.~\ref{figure3}, a single gap $s$-wave model with g$_k$=1 can well reproduce the  experimental data with $\Delta(0)$~=~2.23~$k_BT_c$. This gap magnitude is larger than the value of 1.76~$k_BT_c$ for weakly-coupled BCS superconductors, suggesting moderately strong  coupling in Re$_6$Zr. On the other hand, since as discussed above the samples are near to the dirty limit, a dirty $s$-wave model was also applied, where $\rho_s(T)=\frac{\Delta(T)}{\Delta(0)}\textrm{tanh}[\frac{\Delta(T)}{2k_BT}]$ \cite{Tinkham1996introduction}. The  results are shown by the solid red line, which also well  accounts for the experimental data with $\Delta(0)$~=~2.1~$k_BT_c$. In order to compare with  nodal superconducting scenarios, both a $p$-wave model with g$_k$=sin$\theta$ and a $d$-wave model g$_k$=cos2$\phi$ are also displayed, where $\theta$ is the polar angle and $\phi$ is the azimuthal angle. It is obvious that the superfluid density of $p-$ and $d-$ wave superconductors change significantly with temperature, even at very low temperatures, which is in contrast with the data showing near  temperature independence below around 0.25~$T_c$. Furthermore, we also compare our results with measurements of polycrystalline samples using transverse-field muon-spin rotation ($\mu$SR) measurements from Ref.~\onlinecite{singh2014detection}. These are displayed in Fig.~\ref{figure3} and show that comparable superconducting gap structures are inferred from measurements  of single crystal and polycrystalline samples.

\begin{figure}[t]
\begin{center}
  \includegraphics[width=\columnwidth]{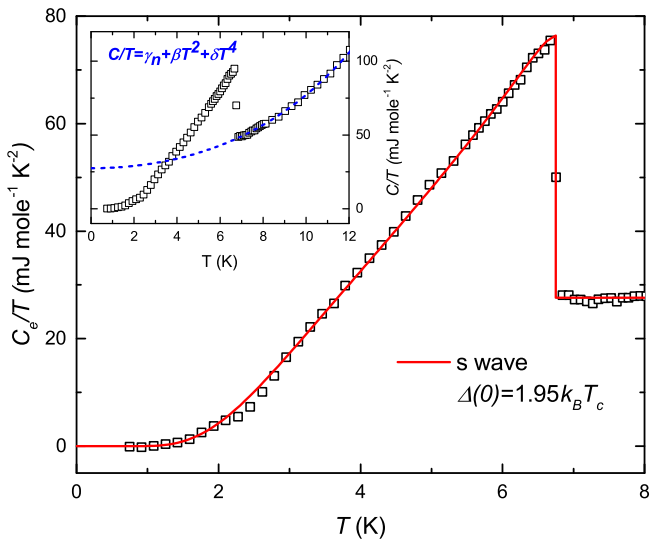}
\end{center}
	\caption{(Color online) Temperature dependence of the electronic specific heat as $C_e(T)/T$ of single crystalline Re$_6$Zr. The solid red line represents the fitting results from a single gap $s$-wave model. The inset displays the total  $C(T)/T$, where the dashed line shows the fit to the normal state.}
   \label{figure4}
\end{figure}

The specific heat measurements of  Re$_6$Zr single crystals were also analyzed, to further characterize the superconducting order parameter. In the inset of Fig.~\ref{figure4}, the total specific heat as $C/T$ is displayed, which contains both electronic and phonon contributions. In the normal state, the data were fitted using $C(T)=\gamma_nT+\beta T^3+\delta T^5$, where $C_e=\gamma_nT$ and $C_{ph}=\beta T^3+\delta T^5$ represent the electron and phonon contributions respectively, and $\gamma_n$ is the Sommerfeld coefficient. The fitting results are shown by the dashed line with  fitted parameters of $\beta$~=~0.39~mJ~mol$^{-1}$~K$^{-4}$, $\delta$~=~1~$\mu$J~mol$^{-1}$~K$^{-6}$, and $\gamma_n$~=~27.4~mJ~mol$^{-1}$~K$^{-2}$, which are consistent with measurements of polycrystalline samples\cite{singh2014detection}.

The main panel of  Fig.~\ref{figure4} displays the temperature dependence of the electronic specific heat as $C_e/T$ in the superconducting state after subtracting the phonon contribution. A sizeable jump is observed around $T_c$ of $\Delta C/\gamma_nT_c$=1.76, which is larger than the value of 1.43 for a  weakly-coupled BCS superconductor, again indicating an enhancement of the coupling strength in Re$_6$Zr. In the superconducting state, the  entropy $S$ can be expressed as\cite{bouquet2001phenmonoligical}:
\begin{equation}
S = -\frac{3\gamma_n}{\pi^3}\int_{0}^{2\pi}\int_{0}^{\infty}[f{\rm{ln}}f+(1-f){\rm{ln}}(1-f)]\rm d\rm{\varepsilon}\rm d\rm{\phi}.
\label{equation3}
\end{equation}

\noindent and therefore the electronic specific heat in the superconducting state can be calculated  as $C_e=T{\rm d}S/{\rm d}T$. Due to $C_e(T)/T$ becoming flat with decreasing temperature for $T~\ll~T_c$, the data were fitted using an isotropic $s$-wave model, and the results are shown in Fig.~\ref{figure4} by the solid red line. It can be clearly seen that this model well describes the experimental data, with a fitted parameter of $\Delta(0)$~=~1.95~$k_BT_c$. This is consistent with the penetration depth and superfluid density results and also indicates a moderately enhanced superconducting gap magnitude.

Since Re$_6$Zr is a NCS, the ASOC will  lift the spin degeneracy of the electron bands, which can potentially give rise to a mixture of spin singlet and spin triplet pairing states, leading to a two-gap structure $\Delta_\pm$~=~$\psi\pm|\mathbf{d}|$, where $\psi$ and $\mathbf{d}$ represent the singlet and triplet components \cite{smidman2017superconductivity}. Therefore the experimental results being consistent with a single nodeless gap implies that the spin triplet component is very small relative to the spin singlet. The origin of TRS breaking in Re$_6$Zr still remains difficult to account for. In LaNiC$_2$, because of the low symmetry of the orthorhombic crystal structure, the breaking of TRS at $T_c$ implies that singlet-triplet mixing is weak, and moreover that the pairing corresponds to a non-unitary triplet state \cite{hillier2009evidence,quintainilla2010relativistic}. Here the lack of a role played by ASOC was corroborated by the breaking of TRS in the similar but centrosymmetric LaNiGa$_2$ \cite{hillier2012nonunitary}, and the observation of two-gap superconductivity in both  LaNiC$_2$ and LaNiGa$_2$ allowed for the proposal of a nodeless even-parity, non-unitary triplet state, where the overall pairing wave function is antisymmetric upon particle exchange due to an antisymmetric orbital index \cite{weng2016two}. Such a scenario is not readily applied to Re$_6$Zr, since  the higher symmetry of the cubic $\alpha$-Mn crystal structure lifts the constraint that TRS breaking at $T_c$ must correspond to non-unitary pairing, and moreover most studies only show evidence for single-gap behavior. As the symmetry analysis of  Re$_6$Zr also allows for mixed singlet-triplet pairing with broken TRS  \cite{singh2014detection}, further studies are highly desirable to understand the origin of TRS breaking and the role of the ASOC in the superconducting state of Re$_6$Zr.\\

\section{Conclusions}

To summarize, we have measured the magnetic penetration depth and electronic specific heat of single crystals of the noncentrosymmetric superconductor Re$_6$Zr. Exponential behavior of $\Delta\lambda(T)$ was observed at $T~\ll~T_c$, which indicates the absence of low energy excitations, and a fully gapped superconducting state. Both the superfluid density and specific heat can be well accounted for by a single gap $s$-wave model across the whole temperature range with a gap magnitude larger than the weak coupling BCS value, providing strong evidence for fully-gapped superconductivity with moderately strong coupling in Re$_6$Zr.

\section{Acknowledgments}

We thank X. Lu for helpful discussions and suggestions.This work was supported by the National Natural Science Foundation of China (No.~11474251, No.~U1632275), the National Key R\&D Program of China (No.~2017YFA0303100, No.~2016YFA0300202),  and the Science Challenge Project of China (No.~TZ2016004).  R.P.S. acknowledges the Science and Engineering Research Board, Government of India for the Young Scientist Grant No. YSS/2015/001799.


\begin{thebibliography}{10}
\bibitem{smidman2017superconductivity}M. Smidman, M. B. Salamon, H. Q. Yuan, and D. F. Agterberg, Superconductivity and spin-orbit coupling in non-centrosymmetric materials: a review. Rep. Prog. Phys. \textbf{80,} 036501 (2017).
\bibitem{BauerNCS} E. Bauer and M. Sigrist, \textit{Non-Centrosymmetric Superconductors: Introduction and Overview}, Lecture Notes in Physics (Springer-Verlag, Berlin, 2012).
\bibitem{gorkov2001superconducting} L. P. Gorkov and E. I. Rashba, Superconducting 2D System with Lifted Spin Degeneracy: Mixed Singlet-Triplet State. Phys. Rev. Lett. \textbf{87,} 037004(2001).
\bibitem{frigeri2004superconductivity}P. A. Frigeri, D. F. Agterberg, A. Koga, and M. Sigrist, Superconductivity without Inversion Symmetry: MnSi versus CePt$_3$Si. Phys. Rev. Lett. \textbf{92,} 097001 (2004).
\bibitem{bonalde2005evidence}I. Bonalde, W. Bramer-Escamilla, and E. Bauer, Evidence for Line Nodes in the Superconducting Energy Gap of Noncentrosymmetric CePt$_3$Si from Magnetic Penetration Depth Measurements. Phys. Rev. Lett. \textbf{94,} 207002 (2005).
\bibitem{yogi2006evidence}M. Yogi, H. Mukuda, Y. Kitaoka, S. Hashimoto, T. Yasuda, R. Settai, T. D. Matsuda, Y. Haga, Y. Onuki, P. Rogl, and E. Bauer, Evidence for Novel Pairing State in Noncentrosymmetric Superconductor CePt$_3$Si: $^{29}$Si-NMR Knight Shift Study. J. Phys. Soc. Jpn. \textbf{75,} 013709 (2006).
\bibitem{Takeya2005} H. Takeya, K. Hirata, K. Yamaura, K. Togano, M. El Massalami, R. Rapp, F. A. Chaves, and B. Ouladdiaf, Low-temperature specific-heat and neutron-diffraction studies on Li$_2$Pd$_3$B and Li$_2$Pt$_3$B superconductors. Phys. Rev. B  \textbf{72,} 104506 (2005).
\bibitem{Nishiyama2005} M. Nishiyama, Y. Inada, and G. -Q. Zheng, Superconductivity of the ternary boride Li$_2$Pd$_3$B probed by $^{11}$B NMR. Phys. Rev. B \textbf{71,}  220505(R) (2005)
\bibitem{yuan2006s}H. Q. Yuan, D. F. Agterberg, N. Hayashi, P. Badica, D. Vandervelde, K. Togano, M. Sigrist and M. B. Salamon, $S$-Wave spin-triplet order in superconductors without inversion symmetry: Li$_2$Pd$_3$B and Li$_2$Pt$_3$B. Phys. Rev. Lett. \textbf{97,} 017006 (2006).
\bibitem{Nishiyama2007} M. Nishiyama, Y. Inada, and G. -Q. Zheng, Spin Triplet Superconducting State due to Broken Inversion Symmetry in Li$_2$Pt$_3$B. Phys. Rev. Lett. \textbf{98,} 047002 (2007).
\bibitem{Takeya2007} H. Takeya, M. ElMassalami, S. Kasahara, and K. Hirata, Specific-heat studies of the spin-orbit interaction in noncentrosymmetric Li$_2$(Pd$_{1-x}$Pt$_x$)$_3$B ($x=0,0.5,1$) superconductors. Phys. Rev. B \textbf{76,}  104506 (2007).
\bibitem{pang2015evidence}G. M. Pang, M. Smidman, W. B. Jiang, J. K. Bao, Z. F. Weng, Y. F. Wang, L. Jiao, J. L. Zhang, G. H. Cao, and H. Q. Yuan, Evidence for nodal superconductivity in quasi-one-dimensional K$_2$Cr$_3$As$_3$. Phys. Rev. B \textbf{91,} 220502(R) (2015).
\bibitem{YPtBinode}     H. Kim, K. Wang, Y. Nakajima, R. Hu, S. Ziemak, P. Syers, L. Wang, H. Hodovanets, J. D. Denlinger, P. M. R. Brydon, D. F. Agterberg, M. A. Tanatar, R. Prozorov. and J. Paglione, Beyond triplet: Unconventional superconductivity in a spin-3/2 topological semimetal. Sci. Adv. \textbf{4,} eaao4513 (2018).
\bibitem{Y2C3NMR}A. Harada, S. Akutagawa, Y. Miyamichi, H. Mukuda, Y. Kitaoka, J. Akimitsu, Multigap Superconductivity in Y$_2$C$_3$: A 13C-NMR Study,  J. Phys. Soc. Jpn. \textbf{76,} 023704 (2007).
\bibitem{Y2C3HC} S. Akutagawa, and J. Akimitsu, Superconductivity of Y$_2$C$_3$ Investigated by Specific Heat Measurement, J. Phys. Soc. Jpn. \textbf{76,} 024713 (2007).
\bibitem{Y2C3MuSR}S. Kuroiwa, Y. Saura, J. Akimitsu, M. Hiraishi, M. Miyazaki, K. H. Satoh, S. Takeshita, and R. Kadono, Multigap Superconductivity in Sesquicarbides La$_2$C$_3$ and Y$_2$C$_3$, Phys. Rev. Lett. \textbf{100,} 097002 (2008).
\bibitem{chen2011evidence}J. Chen, M. B. Salamon, S. Akutagawa, J. Akimitsu, J. Singleton, J. L. Zhang, L. Jiao, and H. Q. Yuan, Evidence of nodal gap structure in the noncentrosymmetric superconductor. Y$_2$C$_3$, Phys. Rev. B \textbf{83,} 144529 (2011).
\bibitem{bauer2009a}E. Bauer, R. T. Khan, H. Michor, E. Royanian, A. Grytsiv, N. Melnychenko-Koblyuk, P. Rogl, D. Reith, R. Podloucky, E. W. Scheidt, W. Wolf, and M. Marsman, BaPtSi$_3$: A noncentrosymmetric BCS-like superconductor, Phys. Rev. B \textbf{80,} 064504 (2009).
\bibitem{Matano2013} K. Matano, S. Maeda, H. Sawaoka, Y. Muro, T. Takabatake, B. Joshi, S. Ramakrishnan, K. Kawashima, J. Akimitsu, and G.~-Q. Zheng, NMR and NQR Studies on Non-centrosymmetric Superconductors Re$_7$B$_3$, LaBiPt, and BiPd,  J. Phys. Soc. Jpn. \textbf{82,} 084711 (2013).
\bibitem{Sun2015} Z. Sun, M. Enayat, A. Maldonado, C. Lithgow, E. Yelland, D. C. Peets, A. Yaresko, A. P. Schnyder, and P. Wahl, Dirac surface states and nature of superconductivity in Noncentrosymmetric BiPd, Nat. Commun. \textbf{6,} 6633 (2015).
\bibitem{jiao2014anisotropic}L. Jiao, J. L. Zhang, Y. Chen, Z. F.Weng, Y. M. Shao, J. Y. Feng, X. Lu, B. Joshi, A. Thamizhavel, S. Ramakrishnan, and H. Q. Yuan, Anisotropic superconductivity in noncentrosymmetric BiPd. Phys. Rev. B \textbf{89,} 060507(R) (2014).
\bibitem{yan2016nodeless}X. B. Yan, Y. Xu, L. P. He, J. K. Dong, H. Cho, D. C. Peets, Je-Geun Park and S Y Li, Nodeless superconductivity in the noncentrosymmetric superconductor BiPd. Supercond. Sci. Technol. \textbf{29,} 065001 (2016).
\bibitem{PbTaSe2SpecH} C. -L. Zhang, Z. Yuan, G. Bian, S. -Y. Xu, X. Zhang, M. Z. Hasan, and S. Jia, Superconducting properties in single crystals of the topological nodal semimetal PbTaSe$_2$.Phys. Rev. B \textbf{93,} 054520 (2016).
\bibitem{pang2016nodeless}G. M. Pang, M. Smidman, L. X. Zhao, Y. F. Wang, Z. F. Weng, L. Q. Che, Y. Chen, X. Lu, G. F. Chen, and H. Q. Yuan, Nodeless superconductivity in noncentrosymmetric PbTaSe$_2$ single crystals. Phys. Rev. B \textbf{93,} 060506 (2016).
\bibitem{wang2016nodeless}M. X. Wang, Y. Xu, L. P. He, J. Zhang, X. C. Hong, P. L. Cai, Z. B. Wang, J. K. Dong, and S. Y. Li, Nodeless superconducting gaps in noncentrosymmetric superconductor PbTaSe$_2$ with topological bulk nodal lines, Phys. Rev. B \textbf{93,} 020503(R) (2016).
\bibitem{barker2015unconventional}J. A. T. Barker, D. Singh, A. Thamizhavel, A. D. Hillier, M. R. Lees, G. Balakrishnan, D McK. Paul, and R. P. Singh, Unconventional Superconductivity in La$_7$Ir$_3$ Revealed by Muon Spin Relaxation: Introducing a New Family of Noncentrosymmetric Superconductor That Breaks Time-Reversal Symmetry. Phys. Rev. Lett. \textbf{115,} 267001 (2015).
\bibitem{singh2014detection}R. P. Singh, A. D. Hillier, B. Mazidian, J. Quintanilla, J. F. Annett, D. M. Paul, G. Balakrishnan, and M. R. Lees, Detection of Time-Reversal Symmetry Breaking in the Noncentrosymmetric Superconductor Re$_6$Zr Using Muon-Spin Spectroscopy. Phys. Rev. Lett. \textbf{112,} 107002 (2014).
\bibitem{matano2016full}K. Matano, R. Yatagai, S. Maeda, and Guo-qing Zheng, Full-gap superconductivity in noncentrosymmetric Re$_6$Zr, Re$_{27}$Zr$_5$, and Re$_{24}$Zr$_5$. Phys. Rev. B \textbf{94,} 214513 (2016).
\bibitem{khan2016complex}M. A. Khan, A. B. Karki, T. Samanta, D. Browne, S. Stadler, I. Vekhter, A. Pandey, P. W. Adams, D. P. Young, S. Teknowijoyo, K. Cho, R. Prozorov, and D. E. Graf, Complex superconductivity in the noncentrosymmetric compound Re$_6$Zr. Phys. Rev. B \textbf{94,} 144515 (2016).
\bibitem{Mayoh2017} D. A. Mayoh, J. A. T. Barker, R. P. Singh, G. Balakrishnan, D. McK. Paul, and M. R. Lees, Superconducting and normal-state properties of the noncentrosymmetric superconductor Re$_6$Zr. Phys. Rev. B \textbf{96,} 064521 (2017).
\bibitem{Bian2016topological}G. Bian, T. -R. Chang, R. Sankar, S. -Y. Xu, H. Zheng, T. Neupert, C. -K. Chiu, S. -M. Huang, G. Q. Chang, I. Belopolski, D. S. Sanchez, M. Neupane, N. Alidoust, C. Liu, B. K. Wang, C. -C. Lee, H. -T. Jeng, C. L. Zhang, Z. J. Yuan, S. Jia, A. Bansil, F. C. Chou, H. Lin, and M. Z. Hasan, Topological nodal-line fermions in spin-orbit metal PbTaSe$_2$. Nat. Commun. \textbf{7,} 10556 (2016).
\bibitem{Guan2016}S. Y. Guan, P. J. Chen, M. W. Chu, R. Sankar, F. Chou, H. T. Jeng, C. S. Chang, and T. M. Chuang, Superconducting topological surface states in the noncentrosymmetric bulk superconductor PbTaSe$_2$. Sci. Adv. \textbf{2,} e1600894 (2016).
\bibitem{Neupane2016observation}M. Neupane, N. Alidoust, M. M. Hosen, J. -X. Zhu, K. Dimitri, S. -Y. Xu, N. Dhakal, R. Sankar, I. Belopolski, D. S. Sanchez, T. -R. Chang, H. -T. Jeng, K. Miyamoto, T. Okuda, H. Lin, A. Bansil, D. Kaczorowski, F. C. Chou, M. Z. Hasan, and T. Durakiewicz, Observation of the spin-polarized surface state in a noncentrosymmetric superconductor BiPd. Nat. Commun. \textbf{7,} 13315 (2016).
\bibitem{luke1998time}G. M. Luke, Y. Fudamoto, K. M. Kojima, M. I. Larkin, J. Merrin, B. Nachumi, Y. J. Uemura, Y. Maeno, Z. Q. Mao, Y. Mori et al., Time-reversal symmetry-breaking superconductivity in Sr$_2$RuO$_4$. Nature (London) \textbf{394,} 558 (1998).
\bibitem{xia2006high}J. Xia, Y. Maeno, P. T. Beyersdorf, M. M. Fejer, and A. Kapitulnik, High resolution polar Kerr effect measurements of Sr$_2$RuO$_4$: evidence for broken time-reversal symmetry in the superconducting state. Phys. Rev. Lett. \textbf{97,} 167002 (2006).
\bibitem{hillier2009evidence}A. D. Hillier, J. Quintanilla, and R. Cywinski, Evidence for Time-Reversal Symmetry Breaking in the Noncentrosymmetric Superconductor LaNiC$_2$. Phys. Rev. Lett. \textbf{102,} 117007 (2009).
\bibitem{quintainilla2010relativistic}J. Quintanilla, A. D. Hillier, J. F. Annett, and R. Cywinski, Relativistic analysis of the pairing symmetry of the noncentrosymmetric superconductor LaNiC$_2$. Phys. Rev. B \textbf{82,} 174511 (2010).
\bibitem{pradnya2017multiband}Pradnya Parab, Deepak Singh, Harries Muthurajan, R. P. Singh, Pratap Raychaudhuri, Sangita Bose, Multiband Superconductivity in the time reversal symmetry broken superconductor Re$_6$Zr. arXiv:1704.06166.
\bibitem{prozorov2000meissner}R. Prozorov, R. W. Giannetta, A. Carrington, and F. M. Araujo-Moreira, Meissner-London state in superconductors of rectangular cross section in a perpendicular magnetic field. Phys. Rev. B \textbf{62,} 115 (2000).
\bibitem{orlando1979critical}T. P. Orlando, E. J. McNiff, S. Foner, and M. R. Beasley, Critical fields, Pauli paramagnetic limiting, and material parameters of Nb$_3$Sn and V$_3$Si. Phys. Rev. B \textbf{19,} 4545(1979).
\bibitem{Goldenfeld1993} P. J. Hirschfeld and N. Goldenfeld, Effect of strong scattering on the low-temperature penetration depth of a $d$-wave superconductor. Phys. Rev. B \textbf{48,} 4219 (1993).
\bibitem{Carrington2003} A. Carrington and F. Manzano, Magnetic penetration depth of MgB$_2$, Physica C \textbf{385,} 205 (2003).
\bibitem{Tinkham1996introduction}M. Tinkham, Introduction to Superconductivity, 2nd (McGraw Hill, New York, 1996).
\bibitem{bouquet2001phenmonoligical}F. Bouquet, Y. Wang, R. A. Fisher, D. G. Hinks, J. D. Jorgensen, A. Junod and N. E. Phillips, Phenomenological two-gap model for the specific heat of MgB$_2$. Europhys. Lett. 56, 856 (2001).
\bibitem{hillier2012nonunitary}A. D. Hillier, J. Quintanilla, B. Mazidian, J. F. Annett, and R. Cywinski, Nonunitary Triplet Pairing in the Centrosymmetric Superconductor LaNiGa$_2$. Phys. Rev. Lett. \textbf{109,} 097001 (2012).
\bibitem{weng2016two}Z. F. Weng, J. L. Zhang, M. Smidman, T. Shang, J. Quintanilla, J. F. Annett, M. Nicklas, G. M. Pang, L. Jiao, W. B. Jiang, Y. Chen, F. Steglich, and H. Q. Yuan, Two-Gap Superconductivity in LaNiGa2 with Nonunitary Triplet Pairing and Even Parity Gap Symmetry. Phys. Rev. Lett. \textbf{117,} 027001 (2016).









\end{thebibliography}
\end{document}